\documentclass[%
 reprint,
superscriptaddress,
 amsmath,amssymb,
pra,
]{revtex4-2}
\usepackage{url}
\usepackage{braket}
\usepackage{graphicx}% Include figure files
\usepackage{dcolumn}% Align table columns on decimal point
\usepackage{bm}% bold math
\usepackage{color,soul}
\usepackage{comment}
\usepackage{url}
\usepackage{siunitx} % for \ohm and SI units
\usepackage[hidelinks,breaklinks=true]{hyperref}
\usepackage{xurl}
\usepackage{newfloat}
\DeclareFloatingEnvironment[name={FIG.}]{appfigure}

\hypersetup{
    colorlinks=true,
    linkcolor=blue,
    filecolor=magenta,      
    urlcolor=cyan,
}
\begin{document}

\preprint{APS/123-QED}

\title{Reproducibility and variability in commercial SiC MOSFETs at deep-cryogenic temperatures
}% Force line breaks with \\

\author{Megan Powell}%
\email{Email: megan.powell@strath.ac.uk}
\affiliation{
Department of Physics, SUPA, University of Strathclyde, Glasgow G4 0NG, United Kingdom}
\author{Euan Parry}%
\affiliation{
Department of Physics, SUPA, University of Strathclyde, Glasgow G4 0NG, United Kingdom}
\author{Conor McGeough}%
\affiliation{
Department of Physics, SUPA, University of Strathclyde, Glasgow G4 0NG, United Kingdom}
\author{Alexander Zotov}%
\affiliation{
Department of Physics, SUPA, University of Strathclyde, Glasgow G4 0NG, United Kingdom}
\author{Alessandro Rossi}
\email{Email: alessandro.rossi@strath.ac.uk}
\affiliation{
Department of Physics, SUPA, University of Strathclyde, Glasgow G4 0NG, United Kingdom}
\affiliation{National Physical Laboratory, Hampton Road, Teddington TW11 0LW, United Kingdom
}

\begin{abstract}
Silicon carbide is a wide-bandgap semiconductor with an emerging CMOS technology platform and it is widely deployed in high power and harsh environment electronics. This material is also attracting interest for quantum technologies through its crystal defects, which can act as spin-based qubits or single-photon sources. In this work, we assess the cryogenic performance of commercial power MOSFETs to evaluate their suitability for CMOS-compatible quantum electronics. We perform a statistical study of threshold voltage and subthreshold swing from 300~K down to 650~mK, focusing on reproducibility and variability. Our results show significant performance degradation at low temperatures, including large gate hysteresis, threshold voltage shifts, and subthreshold swing deterioration. These effects suggest instability in electrostatic control, likely due to carrier freeze-out and high interface trap density, which may pose challenges for the reliable use of this transistor technology towards the realisation of quantum devices or cryo-CMOS electronics.
\end{abstract}

\maketitle

%\tableofcontents
\begin{figure*}[t]
\centering
\includegraphics[scale=0.285]{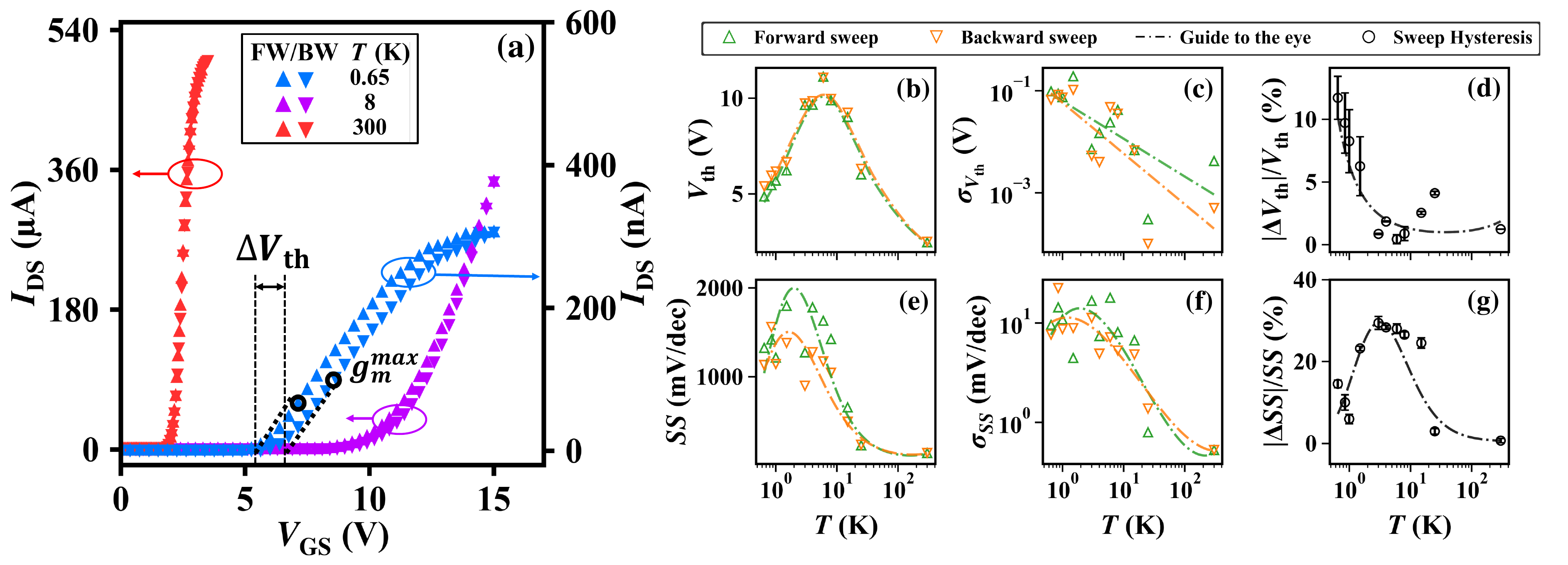}
\caption{Device A characteristics and parameters as a function of temperature. (a) Measured $I_\textup{DS}$-$V_\textup{GS}$ transfer characteristics at $T=0.65$~K, $8$~K, $300$~K (blue, purple and red triangles, respectively). \( V_{\text{DS}}^\textup{0.65~K} \) = 3.8 V, \( V_{\text{DS}}^\textup{8~K} \) = 2.4 V and \( V_{\text{DS}}^\textup{300~K} \) = 0.05 V. FW (upper pointed triangles) and BW (downward pointed triangles) indicate the forward (increasing) and backward (decreasing) voltage sweep direction, respectively. The black circles on the $T$ = 0.65 K dataset indicate the points of maximum transconductance used to extract $V_\textup{th}$ from linear fits, the associated $\Delta V_\textup{th}$ due to hysteresis is also shown.  Parameters extracted from statistical distributions of 50 characteristics as functions of temperature: (b) $V_\textup{th}$, (c) $\sigma_{V_\textup{th}}$, (d) $|\Delta V_{\textup{th}}|/V_\textup{th}$, (e) $SS$, (f) $\sigma_{\textup{SS}}$, (g) $|\Delta SS| /SS$. Those parameters related to $V_\textup{th}$ are quoted at a $V^{eq}_\textup{DS} = 2$~V. The dot-dashed lines in each panel represent guides to the eye to aid evaluation of temperature dependencies. The error bars in (d) and (g) indicate $\pm 1\sigma$ (i.e. one standard deviation) of the relevant distribution.}
\label{fig:stat}
\end{figure*}

\section{\label{sec:Intro}Introduction}

Silicon carbide (SiC) is a wide-bandgap semiconductor extensively deployed in commercial applications, such as high-power and harsh environment electronics~\cite{Kimoto_rev,French2016}. In contrast to other wide-bandgap materials, SiC is the only system that has demonstrated realistic prospects towards the realisation of analog and digital building blocks at integrated circuit level. This has been possible through the development of foundry-based 4H-SiC complementary metal-oxide-semiconductor (CMOS) technology~\cite{romijn2021integrated,Romijn2022}.

SiC has more recently also attracted attention in the field of quantum technologies~\cite{Awschalom2018}. Specifically, intrinsic or implanted defects in the SiC crystal can function as spin-based quantum bits (qubits)~\cite{son_rev} or environmental sensors~\cite{abraham2021} and exhibit spin-dependent photonic emission, offering potential for both quantum computing and networking~\cite{Lohrmann_2017}.
Current methods for addressing quantum states in SiC rely on optical scanning techniques on barely processed wafers, an approach not directly amenable to integration or mass production. Exploiting the existing SiC CMOS transistor  technology towards quantum state control and readout could pave the way for industry-compatible integrated quantum electronics, an approach that is already gaining momentum in other CMOS-compatible materials~\cite{banerjee2024materialsquantumtechnologiesroadmap}, primely in silicon~\cite{mfgz21}.

A critical step in validating SiC MOS systems as a platform for integrated quantum technologies is the evaluation of their electronic properties at cryogenic temperatures. Low-temperature operation is often indispensable for preserving fragile quantum attributes against thermal fluctuations. For example, spin state readout  systems based on electrometry~\cite{elze}, must be operated at $4~$K or below to achieve good sensitivity via charge quantisation. Understanding SiC metal oxide semiconductor field effect transistors (MOSFETs) behaviour at such temperatures may therefore be important for envisaging analogous electrical readout schemes~\cite{Burkard_Rev} or designing cryo-CMOS control electronics~\cite{eastoe2024method, patra2017cryo}. To this aim, some desirable features should be high-quality ohmic contacts (linear IV with contact resistance no greater than 1 k$\Omega$ \cite{Angus2007}) for ease of control of the charge reservoirs, sharp channel gating and/or pinch-off to fine tune confinement potentials (at the level of 8 mV/dec at cryogenic temperature~\cite{roche2012tunable}) , as well as overall reproducibility and stability of transport characteristics ($1$\% or less parameter spread with repeated or hysteretic measurements \cite{eastoe2024method}). 

In this work, we characterise commercial vertical SiC power MOSFETs at cryogenic temperatures with an eye to ascertain whether current commercial devices and processes already hold prospect for the realisation of integrated quantum electronics. We carry out a statistical study using repeated IV measurements of two devices at different temperatures to extract threshold voltage and subthreshold swing, considered to be key metrics to assess reproducibility and stability. We study two nominally identical commercial transistors in a wide temperature range from $300~$K down to $650~$mK. We observe a systematic degradation of device performance, as the temperature of operation is decreased, and discuss possible physical origins. Although such cryogenic behaviour appears to be similar for both investigated samples, an increased inter-device variability with decreasing temperature is also reported.

%We suggest that a combination of carrier freeze-out and high trap density are responsible for increased levels of gate voltage hysteresis, long-term drift and charge instability.    

\section{\label{sec:Methods}Methods}

The devices under test are two nominally identical bare die n-channel vertical $1.2$~kV 4H-SiC MOSFETs developed by Wolfspeed (CPM3-1200-0013A) with typical threshold voltage of $2$~V and ON resistance of $13$~m$\Omega$ at room temperature.
%Any dimensions of the device are not disclosed by the manufacturer. Additionally, the device consists of a body diode between source and drain contacts.
For convenience, they will be named \textit{Device A} and \textit{Device B}. These devices are vertical diffusion MOSFETs with source (S) and gate (G) metal contacts located at the top of the chip, whilst the drain (D) contact occupies the entire flip side of the chip. Due to often analogous results between the two samples, on occasions only one device's analysis is presented in the main text (see appendix for comprehensive datasets). 

We measure transport characteristics at multiple temperatures by thermal anchoring the sample holder containing each bond-wired chip onto the mixing chamber plate of a dilution refrigerator. The plate temperature ($T$) can be controlled with dedicated heating devices and we set $0.65$~K~$\leq T\leq25$~K. During the experiments, temperatures were increased in sequential order starting from the lowest. Even though the refrigerator is capable of reaching temperatures as low as tens of millikelvin, the device power dissipation during operation generates sufficient heat to exceed the cooling power of the cryogenic system (approximately $300~\mu$W at $0.10~$K). We point out that the main source of heat in this case originates from transient currents in the loom wires forming an RC circuit with the gate electrode during repeated $V_\textup{GS}$ sweeps, as opposed to a modest steady state dc power. This has prevented us from achieving a stable temperature below approximately $0.65$~K without slowing exceedingly the measurements down due to long RC time constants. Note that operation at these slightly raised temperatures has the advantage that the thermometer temperature closely tracks the effective carrier temperature in the device, whereas at lower temperatures this would be unreliable due to poor electron-phonon coupling~\cite{giazo} .\\\indent

Two separate source-measure units (SMUs) are employed for acquiring IV traces. One SMU is used across drain-source contacts to apply bias voltage ($V_\textup{DS}$), and
measure current ($I_\textup{DS}$). The other SMU is used to apply
gate voltage ($V_\textup{GS}$), whilst monitoring that the leakage
current ($I_\textup{GS}$) remains within nominal values of operation set by the manufacturer, i.e. $I_\textup{GS} \leq$ 30 nA. The
source contact is always kept at reference ground defined
by the Low terminal of an SMU. All output and transfer
characteristics are acquired in Kelvin (four wire) mode to minimise
voltage drop contributions from the set-up wiring. To robustly investigate each performance metric, we have built
statistically relevant datasets by acquiring the same IV
trace of a single device at least 50 times per temperature (see appendix).
This has enabled us to record not only the temperature
dependence of each parameter of interest, but also how
their repeatability and stability were affected by temperature and voltage sweep direction.

%To be able to quote a single parameter value per temperature with a common $V_\textup{DS}$, we corrected to $V_\textup{DS} = 2 V$ using two different methods. 
%Firstly to correct $V_\textup{th}$, the effect of $V_\textup{th}$ against $V_\textup{DS}$ was examined at 300 K - a monotonic increasing relationship was apparent. A least squares regression line was fitted to the data, and the gradient of the line is given by:
%\begin{equation}\label{eq:th_correction}
%\begin{split}
%grad &= \frac{\Delta V_{TH}}{\Delta V_{DS}}
 %   = \frac{V_{TH(\mathrm{M})} - V_{TH(\mathrm{C})}}
 %          {V_{DS(\mathrm{A})} - V_{DS(\mathrm{C})}}\\
%V_{TH(\mathrm{C})}
%    &= V_{TH(\mathrm{M})} 
%       - m\,\bigl[V_{DS(\mathrm{A})} - V_{DS(\mathrm{C})}\bigr],
%\end{split}
%\end{equation}

%where M, C and A stand for Measured, Corrected and Applied respectively. Secondly, to correct $SS$, another linear fit was undertaken to an $SS$ against $V_\textup{DS}$ plot per temperature. Once the equation for this line was extracted, finding the intersect and gradient - we could apply this to our $V_\textup{DS}$ of choice and find $SS$.

\section{\label{sec:Results}Results}

In this section, we present the experimental results, beginning with an overview of the device performance through analysis of the MOSFET transfer characteristics and the key parameters extracted from the corresponding IV curves. We then discuss the emergence of temperature-dependent Schottky behavior at the source and drain contacts. Finally, we describe time-dependent effects observed at cryogenic temperatures and the resulting development of a training protocol aimed at achieving more stable and repeatable device characteristics.

\subsection{Device Performance}

Figure~\ref{fig:stat}(a) shows $I_\textup{DS}$-$V_\textup{GS}$ transfer characteristics at three representative temperatures (see additional traces in Appendix). From these characteristics in linear and log scale, we extract the device's threshold voltage ($V_\textup{th}$) and the subthreshold swing ($SS$) respectively. $V_\textup{th}$ is obtained as the zero current intercept by a linear extrapolation from the point of maximum transconductance ($g^\textup{max}_\textup{m}$)~\cite{sze}, as shown in the main panel. Note that each transfer characteristics is measured at a different $V_\textup{DS}$, depending on the temperature of operation. This stems from the fact that S/D contacts develop an increasingly severe Schottky-type behaviour as temperature decreases and, therefore, we need to increase $V_\textup{DS}$ accordingly to attain transport through the transistors~\cite{gammon2013modelling}. To account for the effect of different bias and ensure fairness in the extraction of $V_\textup{th} (T)$, we quote all threshold voltages at an equivalent bias $V^{eq}_\textup{DS}=2$~V, by using an extrapolation technique discussed in Section~\ref{sec:ResultsSch}.\\\indent
$SS$ is obtained by calculating the inverse gradient of the log-linear characteristics in the weak inversion regime achieved at two fixed values of current ($I_{DS,1}=2~$nA; $I_{DS,2}=10~$nA) using~\cite{sze, neamen}:
\begin{equation}
\label{eq:ss}
\mathrm{SS}
= \frac{dV_{GS}}{dI_{DS}}
= \frac{V_{\mathrm{GS}}\,(I_{DS,2})
        - 
        V_{\mathrm{GS}}\,(I_{DS,1})}
       {\log_{10}(I_{DS,2}) \;-\; \log_{10}(I_{DS,1})}
\end{equation} 
as illustrated in Appendix (see Fig.~\ref{fig:AllTempTransfer}). The use of two values of fixed current allows us to neglect differing $V_\textup{DS}$ on the extraction of $SS$~\cite{yoshi}. The results of the statistical study of device parameters as functions of temperature  are shown in Fig.~\ref{fig:stat}(b)-(g). Figure~\ref{fig:stat}(b) illustrates the  dependence of $V_\textup{th}$ on temperature. One can see that with decreasing temperature from ambient conditions, it increases to a maximum value of $V_\textup{th}\approx 11$~V achieved at $T\approx 8$~K and then steadilydrops at deeper cryogenic temperatures. At the higher end of the temperature range, there is a very good agreement between data measured by increasing (forward mode, FW) and decreasing (backward mode, BW) $V_\textup{GS}$. This agreement progressively deteriorates as the temperature decreases, particularly to the left of the voltage peak ($T\leq 8$~K). We suggest that the non-monotonic behaviour of $V_\textup{th} (T)$ can be ascribed to competing temperature effects in intrinsic carrier concentration, acceptor freeze-out and interface trap occupancy. In particular, it has been observed before in similar devices that $V_\textup{th}$ increases with decreasing temperature because interface traps become increasingly negatively charged and intrinsic carrier density reduces~\cite{MATOCHA20081631, chen2015}. It is plausible that these two effects saturate for sufficiently low temperature, say around $8~$K in our case. For lower $T$, freeze-out in the p-type channel region may result in a sharp decrease in the Fermi surface potential and, consequently, in the threshold voltage. The temperature at which the crossover between different dominant effects occurs is likely dependent on device doping profile and density of interface traps. In fact, for other SiC MOSFET technologies this has been observed at higher temperatures~\cite{gui2018characterization, woodend} and can be correlated to the high interface trap density of the SiC/SiO$_2$ interface~\cite{pascu2020ultrashallow}.
\begin{figure}[b]
  \centering
  \includegraphics[width=1.0\linewidth]{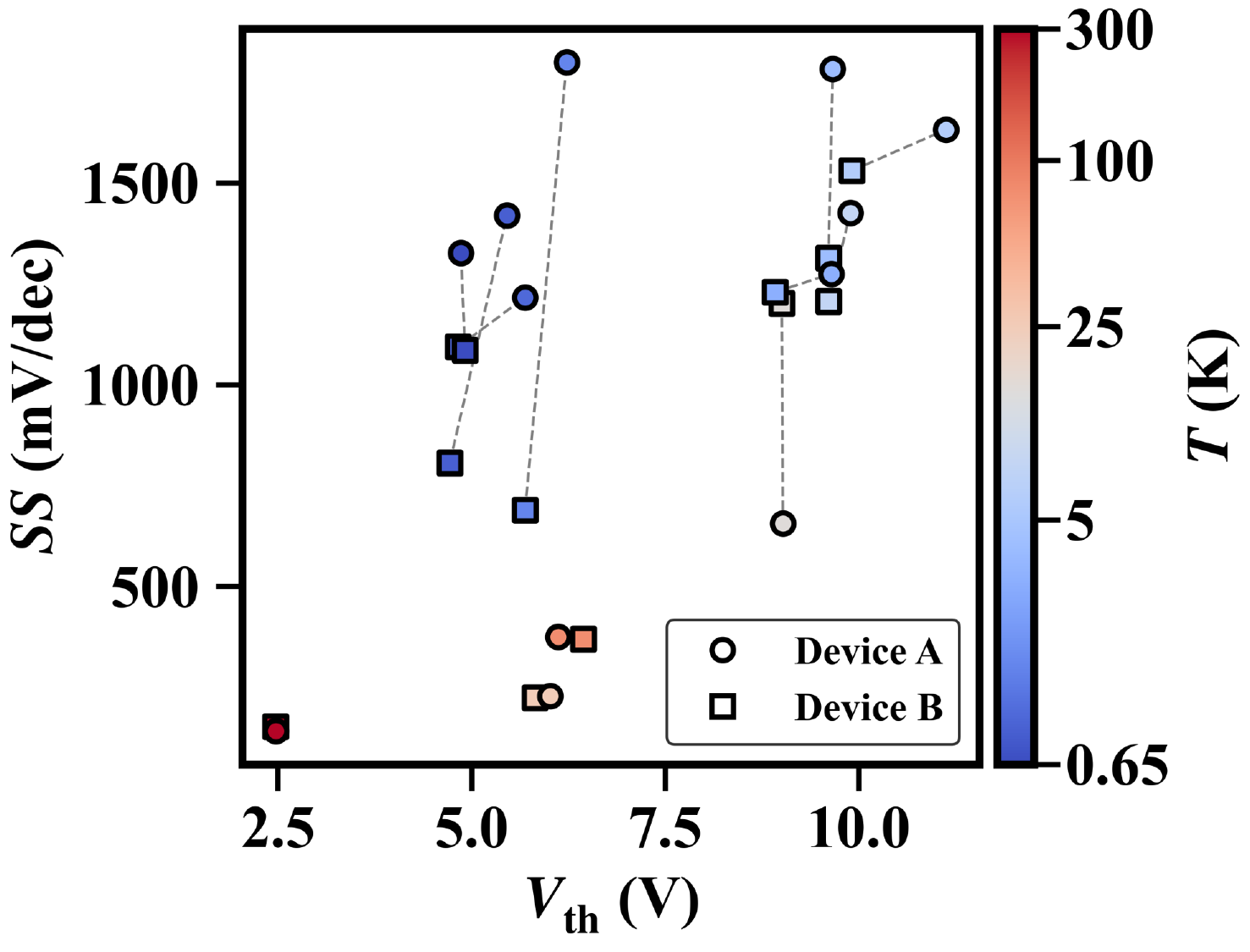}
  \caption{Comparison of performance metrics ($SS$, $V_\textup{th}$) between Device A (circles) and Device B (squares) as a function of temperature (color scale). Dashed lines are guides to the eye highlighting inter-device deviation at equal temperatures. Each data point value is extracted as the average from 50 repeat characteristics, as in Figure 1.} 
  \label{fig:DeviceComparison}
\end{figure}
Figure~\ref{fig:stat}(c) illustrates the statistical spread of $V_\textup{th} (T)$ (standard deviation $\sigma_{V_\textup{th}}$), as obtained from the Gaussian fit to the histogram built with 50 nominally identical IV runs (see appendix). One can see an overall increase of distribution spread for decreasing temperature, which indicates that the devices are subject to increasing instabilities at cryogenic conditions. Interestingly, at the highest temperatures, the instabilities appear to be more marked for the FW measurement mode, whereas at the lowest end of the range the distinction between FW and BW mode tends to fade away. In Fig.~\ref{fig:stat}(d), we measure the effect of gate voltage hysteresis on $V_\textup{th} (T)$. Specifically, we calculate $\Delta V_\textup{th} = V^{BW}_\textup{th} - V^{FW}_\textup{th}$ at each temperature of interest. The plot shows that the relative shift in threshold is quite small (few percent) and approximately constant from room temperature down to $T\approx 1.5$~K. At deeper cryogenic temperatures,  a steady increase in hysteretic effects is observed. This is also accompanied by a stark increase in statistical spread (see error bars). We speculate that threshold voltage hysteresis and instability go hand in hand and may both originate from the large shift introduced by increasingly negatively charged interface traps~\cite{MATOCHA20081631}.

In Fig.~\ref{fig:stat}(e)-(g), a statistical analysis is reported for $SS$, calculated using Equation~\ref{eq:ss}. Contrary to theoretical predictions for MOSFETs by which $SS$ should steadily decrease with decreasing temperature~\cite{beckers2019theoretical}, we see a significant growth down to approximately $4~$K followed by a decrease at lower temperatures, as indicated in panel (e). This is accompanied by a similar trend in statistical spread shown in panel (f). By calculating $\Delta SS = SS^{FW} - SS^{BW}$, one can conclude that the relative hysteresis is small at room temperature, but it peaks at about $30$\% at $4~$K and substantially decreases at lower temperatures, as reported in Fig.~\ref{fig:stat}(g). The rapid increase in $SS$ with decreasing temperature is once again consistent with an increase in the density of SiC/SiO$_2$ interface traps, which ultimately degrades the transport characteristics. In fact, it has been previously seen that higher interface trap density results in $SS$ degradation in SiC~\cite{yoshi} and Si \cite{oka2023milli}. Note that, it appears that the performance slightly improves at deep cryogenic temperatures even though the room temperature situation is not fully restored at the lowest temperature of our experiments ($0.65~$K).  \\\indent
We now turn to discuss the effect of temperature on inter-device variability. In Fig.~\ref{fig:DeviceComparison}, a comparison between the performance of Device A and B is presented as a function of temperature. One can notice that in the upper temperature range (roughly $15~$K$<T<300~$K) the agreement between device metrics is quite good, indicated by a modest separation between isothermal data points (dashed lines). However, as the temperature is further decreased this separation markedly increases. This is generally consistent with the observed degradation of device stability, which is likely contributing towards a more severe inter-device variability at deep cryogenic temperature. We point out that given the limited number of samples investigated for this inter-device study, firmer statistical conclusions can only be drawn after a larger sample pool will be examined.

\begin{figure}[ht]
  \centering
  \includegraphics[width=1.05\linewidth]{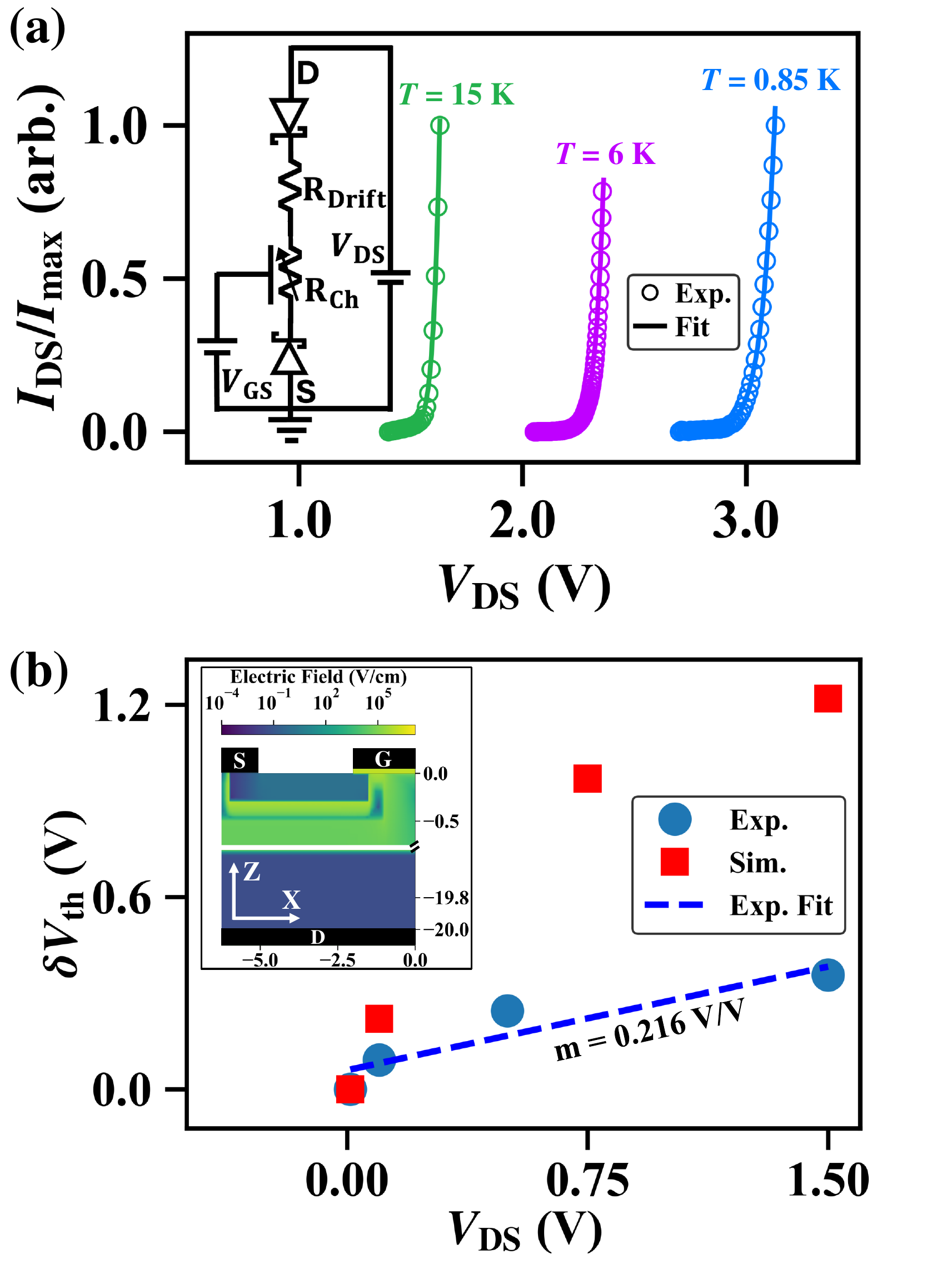}
  \caption{(a) Device A normalised $I_\textup{DS}$-$V_\textup{DS}$ experimental curves and fits at $T=15$~K (green), $T=6$~K (purple) and $T=0.85$~K (blue). \( V_{\text{GS}}^\textup{0.85~K} \) = 8 V, \( V_{\text{GS}}^\textup{6~K} \) = 10 V and \( V_{\text{GS}}^\textup{15~K} \) = 8 V are chosen to be greater than $V_\textup{th}$ at the relevant temperature, so that the transistor is in the ON state. Insert depicts an equivalent circuit diagram of the device presenting the S/D contacts as Schottky diodes at cryogenic temperatures. (b) $V_\textup{th}$ shift as a function of $V_\textup{DS}$ from experiments (circles) and TCAD simulations (squares) at $T = 300~$K. The dashed line indicates a linear fit to the experimental data used to obtain the correction factor  employed to extrapolate $V_\textup{th}(T)$ at an equivalent $V_\textup{DS}=2~$V for all temperatures (as reported in Fig.~\ref{fig:stat}). The insert shows a representative electric field map in a vertical transistor of similar dimensions as the one measured, as obtained from TCAD simulations. The axis labels are in $\mu$m. Simulation is carried out at $V_\textup{DS}=1.5~$V, $V_\textup{GS}=10~$V.}
  \label{fig:sch}
\end{figure}

\subsection{\label{sec:ResultsSch}Cryogenic Schottky Contacts}

As mentioned earlier, we observe increasingly non-linear S/D contact characteristics as temperature decreases. Figure~\ref{fig:sch}(a) reports the normalised $I_\textup{DS}$ as a function of $V_\textup{DS}$ at three representative cryogenic temperatures. One can see a typical non-linear Schottky diode dependence with increasing turn-on voltages for decreasing temperature. Note that at room temperature the linear IV relationship expected for ohmic contacts is observed (not shown). Differing V$_\textup{GS}$ values are used to ensure the transistor's channel is consistently in full inversion whatever the temperature of operation (V$_\textup{th}$ is temperature dependent as discussed earlier).

We argue that, as temperature is reduced, carrier freeze-out in the $n++$ contact regions makes tunneling through the metal-semiconductor Schottky barrier increasingly unfavorable due to an increase in barrier width ($d$)~\cite{neamen}. Hence, the source and drain contacts, acting as Schottky diodes at cryogenic temperature, give rise to the equivalent circuit diagram shown in the inset of Fig.~\ref{fig:sch}(a). The main circuit elements are a small channel resistance ($R_\textup{Ch}$) achieved by choosing $V_\textup{GS}>>V_\textup{th}$, a freeze-out dependent resistance of the transistor's drift region ($R_\textup{Drift}$) and two back-to-back Schottky diodes one for each of S and D contacts. We consider that the dominant component limiting the current is the S diode which operates in reverse bias mode, i.e. until this achieves reverse tunneling breakdown, only a limited current flows through the transistor. We model this scenario by assuming that thermionic emission is suppressed and fit the data using a Fowler-Nordheim tunneling model~\cite{sze}:

\begin{equation}\label{eq:FN_EQ}
I(V_{\!DS})
\;=\;
A\;
\frac{\alpha}{\phi}
\left [\frac{V_{\textup{DS}}-V_{0}}{d}\right ]^{2}\
\exp\!\Biggl(-\,\beta\;\frac{d\phi^{3/2}}{{V_{\textup{DS}} - V_{0}}}\Biggr)
\end{equation}

%\begin{equation}
%\alpha \;=\; \frac{e^3}{16\,\pi^2\,\hbar},
%\end{equation}

%\begin{equation}
%\beta \;=\; \frac{4\,\sqrt{2\,m_{e}}}{3\,e\,\hbar},
%\end{equation}
where $\alpha$ and $\beta$ are fixed terms that can be calculated directly from constants of nature and the effective electron mass ($m^* = 0.42m_e$ \cite{son1995electron}), $A=60~$mm$^{2}$ is the estimated contact size of the S metal pad, $\phi$ is the Schottky barrier height, which we keep constant ($\phi$ = 0.9 eV~\cite{koliakoudakis2008cr}) because only mildly temperature dependent compared to other parameters. There are two fitting parameters: $d$ and $V_\textup{0}$ (the voltage drop across temperature dependent $R_\textup{Drift}$). From the fits of several cryogenic IV curves, we see that both $d$ and $V_\textup{0}$ increase with decreasing temperatures (see table in appendix). This indicates that at lower temperatures the tunnel barrier becomes increasingly thick and the drift region becomes more resistive. These findings are qualitatively consistent with the possible effects of carrier freeze-out. In fact, one expects an inverse quadratic relationship between the barrier width and the effective carrier density in the semiconductor side of a metal-semiconductor junction and an exponential decay of ionization-activated carriers~\cite{sze, neamen}.
\\\indent
The need for relatively large and variable $V_\textup{DS}$ values to achieve transport at low temperatures may affect our ability to consistently extract $V_\textup{th}(T)$. In fact, we observe a clear shift in $V_\textup{th}$ ($\delta V_\textup{th}$) when $V_\textup{DS}$ is changed, as shown in Fig.~\ref{fig:sch}(b). In planar MOSFETs this is typically attributed to drain-induced barrier lowering, a well-established consequence of short-channel effects~\cite{sze}. However, in our case the application of positive drain voltage has the opposite effect, as it operates in competition with the gate voltage by making it harder to invert the channel, leading to a positive voltage gradient (see dashed line). We assume that this effect is to first approximation temperature independent because related to device electrostatics alone. Hence, we use the room temperature experimental gradient shown in Fig.~\ref{fig:sch}(b) as a correction factor ($m$)  to quote all $V_\textup{th}$(T) obtained from the statistical studies at an equivalent $V_\textup{DS}=2~$V.\\\indent
To explain this electrostatic effect, we argue that it stems from the vertical architecture of the transistor which results in large $V_\textup{DS}$ having a gating effect in competition with the top gate electrode and it is akin to dynamic $V_\textup{th}$ tuning via back gating used in some advanced CMOS nodes (e.g. fully-depleted silicon-on-insulator). To corroborate this assumption, we run Sentaurus TCAD simulations and study the electric field profile in the vertical transistor channel as a function of $V_\textup{DS}$, see insert of Fig.~\ref{fig:sch}(b). Similar to the experimental case, the threshold voltages extracted from the simulations show a positive gradient, suggesting that the electrostatic effect of the drain electrode can be assimilated to a back gate. We acknowledge that in Fig.~\ref{fig:sch}(b) the agreement between experimental and simulated $\delta V_\textup{th}$ is not excellent probably because the device physical dimensions used for the simulations do not match the real transistor size, which is not fully disclosed by the manufacturer. Nonetheless, the evidence of a drain-induced gating effect is robust, originating from electrostatic considerations alone. 
\begin{figure}[t]
  \centering
  \includegraphics[width=0.99\linewidth]{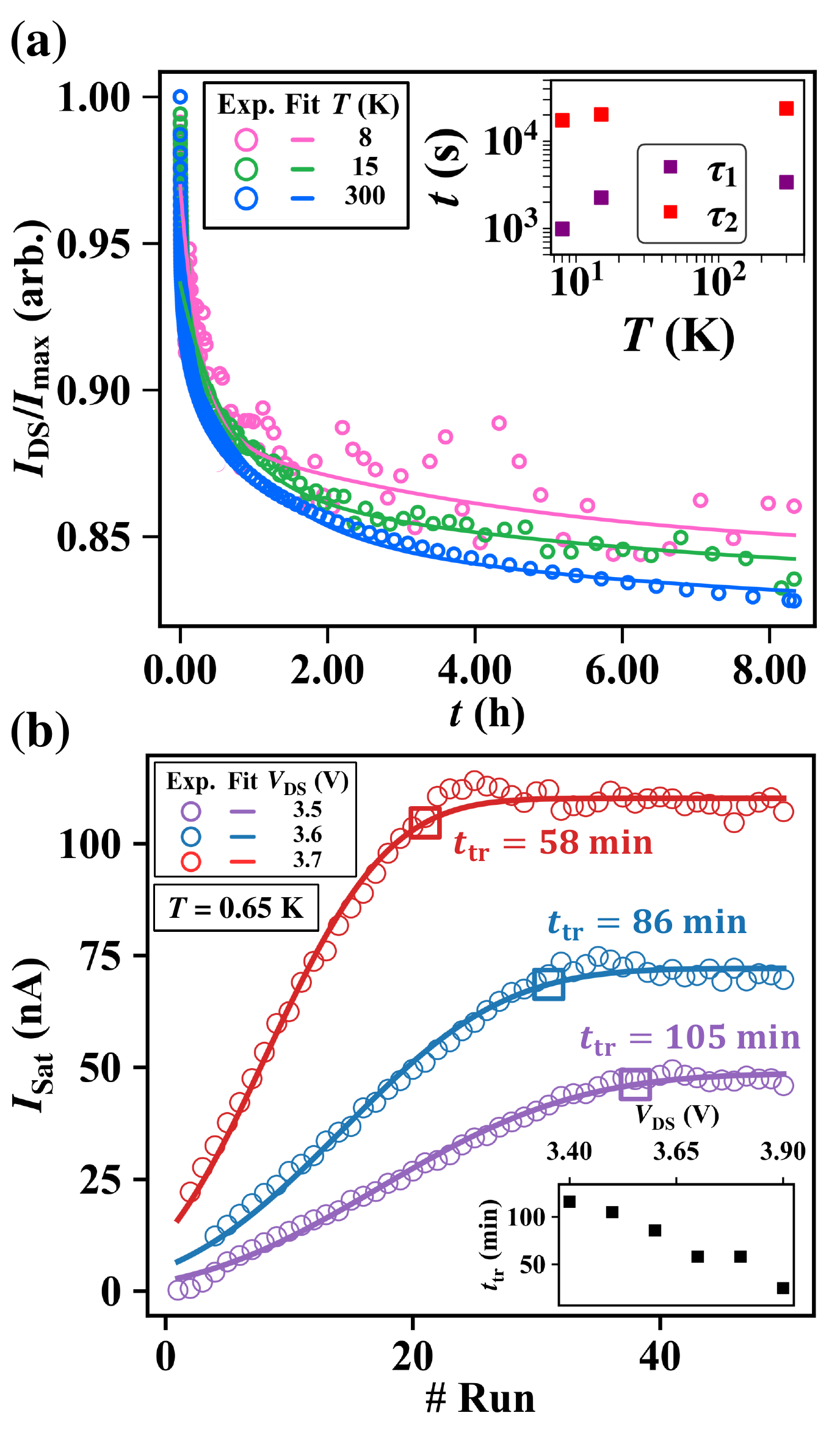}
  \caption{ (a) Device A's experimental $I_\textup{DS}$ drift (circles) and double exponential decay fits (solid lines) as a function of time at $T=8~$K (pink), $T=15~$K (green) and $T=300~$K (blue). The insert depicts the extracted time constants as a function of temperature, note the uncertainty in $\tau$ is shown in Table~\ref{tab:tau_table} of the Appendix. (b) Device B's training protocol at $0.65$~K for $V_\textup{DS}=3.5~$V (purple), $V_\textup{DS}=3.6~$V (blue), and $V_\textup{DS}=3.7~$V (red). Experimental $I_\textup{Sat}$ (circles) and fits (solid lines) as a function of runs. The fits are exponential functions with single time constants reaching 95\% of the saturation current at the points indicated by squares. The insert depicts $t_\textup{tr}$ vs $V_\textup{DS}$ as extracted from the fits (squares in the main panel).}
  \label{fig:time}
\end{figure}

\subsection{Device Drift and Training}

We now turn to discuss significant instability in the device characteristics, which rendered necessary the establishment of a device training protocol prior to collection of the statistical datasets discussed so far. 
In Fig.~\ref{fig:time}(a), we show a significant $I_\textup{DS}$ drift when the device is left idle at the $g_\textup{m}^\textup{max}$ operation point. For all three representative temperatures examined, we observe an initial rapid current drop followed by a slower leveling off. We argue that this behaviour is consistent with the dynamics of oxide-trap charging often reported for SiC MOSFETs~\cite{lelis2014basic}. To model this effect, we fit the experimental data to a double exponential decay function and extract two time constants relevant to the initial fast current drop ($\tau_1$) and the slower saturation ($\tau_2$). As shown in the insert of Fig.~\ref{fig:time}(a), although the two constants differ by approximately an order of magnitude, both decrease with decreasing temperature (see also Table~\ref{tab:tau_table} in Appendix to confirm that the uncertainty of each time constant is small compared to the magnitude of the temperature dependence). This suggests that both processes responsible for the drift are increasingly fast in reaching steady states of operation the lower the temperature. Our observations are qualitatively in line with two oxide-trap charging mechanisms, one occurring rapidly because located at the semiconductor-oxide interface, and another one happening more slowly because located deeper into the oxide layer~\cite{LELIS201832}. Both processes can be accelerated by reducing random thermal fluctuations, which, in turn, makes the gate electrostatics more efficient in attracting and trapping electrons at the oxide.\\\indent
As shown in Fig.~\ref{fig:time}(a), it would have taken an impractically long time to wait for the current to reach a saturation point before collecting statistics. Hence, we implemented a protocol to accelerate the processes of interface and oxide trapping, which we refer to as device training (also known as gate-bias stressing)~\cite{shen2022time}. In Fig.~\ref{fig:time}(b), we plot the maximum value of current reached during a training run ($I_\textup{Sat}$) as a function of the number of consecutive runs performed at $T=0.65$~K. Each run lasts approximately $165$~s and is defined as a forward gate sweep followed by a reverse gate sweep which take the transistor from the OFF state to the ON saturation state and back. It is evident that through the training procedure the transport characteristics undergoes an initial ramp-up stage followed by a leveling off of the maximum current value after a critical amount of time, which we call the cumulative training time ($t_\textup{tr}$). For $t> t_\textup{tr}$, no further dependence of the IV characteristics on measurement run is observed, which indicates that our parameter extraction method is not affected by device degradation (see also Fig.~\ref{fig:Param_VS_Repeat} in Appendix). As depicted in the inset of Fig.~\ref{fig:time}(b),  $t_\textup{tr}$ depends on $V_\textup{DS}$ and can be reduced by running the protocol at higher voltages. We apply this training protocol prior to the acquisition of IV characteristics every time the experimental temperature was modified at the lower end of the temperature range ($T\leq15~$K). At higher temperatures, the requirement for device training is less stringent. %\textcolor{red}{Moreover, Fig.~\ref{fig:Param_VS_Repeat} in the Appendix is shown to evidence that deviation in device parameters is not a result of this degradation with time which is eliminated with device training.}
\section{\label{sec:Disc}Conclusion}
 
We have presented a statistical study of commercially available SiC power MOSFETs at deep cryogenic temperatures, to assess the viability of existing device technology for quantum applications. Our results suggest that the device technology examined here may not be suitable for this purpose, as device performance, reliability and variability all deteriorate at low temperature. 
When benchmarking our results against typical or early versions of silicon quantum devices, several key limitations become evident. The deterioration of device performance at reduced temperatures leads to significant setbacks. Whereas S/D contacts in functional quantum devices typically exhibit linear IV characteristics and contact resistances below 1 k$\Omega$ \cite{Angus2007} at cryogenic temperatures, our devices instead display Schottky-like behaviour, with even the linear-region resistance reaching approximately 500 k$\Omega$. The lack of high-quality ohmic contacts at low temperature would adversely impact the prospects of using these devices for quantum state readout. This stems from the need for  fast readout electronics compatible with single-shot charge detection~\cite{elze}, which would be difficult to implement with highly resistive contacts. %Furthermore, for a $V^{eq}_\textup{DS}=2$~V, the observed device variability shows a spread roughly twice that reported in the literature ($\approx$ 2\% vs. 1\%) \cite{eastoe2024method}. 
We also report a pronounced degradation in $SS$ which would severely constrain the ability to fine-tune confinement gates, requiring voltages hundreds of times larger than those in other semiconductor quantum devices \cite{roche2012tunable}. This suggests that the ability of modulating the channel potential is weak and negatively affected by interface traps or leakages leading to poor control of tunnel barriers~\cite{Angus2007}. Additionally, the observation of significant voltage hysteresis (well above the functional limit of $1$\% reported for silicon), especially at deep cryogenic temperatures, indicates unstable and history-dependent electrostatic control. This behaviour could limit the ability to form and tune quantum dots with repeatable and predictable charge occupancy, essential ingredients for both MOS readout electronics and qubit devices~\cite{Burkard_Rev}.  \\\indent
Despite these limitations, our findings also highlight specific technological targets that could make SiC a more viable platform for cryogenic quantum electronics. The main challenges we observe are primarily rooted in material science issues, such as interface trap density, oxide quality, and contact formation. Encouragingly, these are all areas where continued progress is being made within the SiC power device community. Beyond material-level improvements, further gains could be achieved through tailored device design choices, made possible by the increasingly mature CMOS technology available for SiC.

\begin{acknowledgments}
The authors thank V. Shah  for useful discussions, N. Owen for TCAD advice, as well as J. Gillan and A. Robbins for technical support. A.R. acknowledges support from the UKRI Future Leaders Fellowship Scheme (Grant agreement: MR/T041110/1). 
\end{acknowledgments}

\section*{Data availability statement}

\begingroup
\sloppy
\emergencystretch=3em
The data that support the findings of this study are openly available at the following URL/DOI:
\url{https://pureportal.strath.ac.uk/en/datasets/data-for-reproducibility-and-variability-in-commercial-sic-mosfet/}
\cite{powell2026data}.
\par
\endgroup

\newpage

\bibliography{ref_main}

@PREAMBLE{
 "\providecommand{\noopsort}[1]{}" 
 # "\providecommand{\singleletter}[1]{#1}%" 
}

@article{giazo,
  title = {Opportunities for mesoscopics in thermometry and refrigeration: Physics and applications},
  author = {Giazotto, Francesco and Heikkil\"a, Tero T. and Luukanen, Arttu and Savin, Alexander M. and Pekola, Jukka P.},
  journal = {Rev. Mod. Phys.},
  volume = {78},
  issue = {1},
  pages = {217--274},
  numpages = {0},
  year = {2006},
  month = {Mar},
  publisher = {American Physical Society},
  doi = {10.1103/RevModPhys.78.217},
  url = {https://link.aps.org/doi/10.1103/RevModPhys.78.217}
}

@article{Kimoto_rev,
doi = {10.7567/JJAP.54.040103},
url = {https://dx.doi.org/10.7567/JJAP.54.040103},
year = {2015},
month = {mar},
publisher = {The Japan Society of Applied Physics},
volume = {54},
number = {4},
pages = {040103},
author = {Kimoto, Tsunenobu},
title = {{M}aterial science and device physics in {S}i{C} technology for high-voltage power devices},
journal = {Japanese Journal of Applied Physics}
}

@Article{French2016,
author={French, Paddy
and Krijnen, Gijs
and Roozeboom, Fred},
title={{P}recision in harsh environments},
journal={Microsystems {\&} Nanoengineering},
year={2016},
month={Oct},
day={10},
volume={2},
number={1},
pages={16048},
doi={10.1038/micronano.2016.48},
url={https://doi.org/10.1038/micronano.2016.48}
}

@article{romijn2021integrated,
  title={{I}ntegrated digital and analog circuit blocks in a scalable silicon carbide {CMOS} technology},
  author={Romijn, Joost and Vollebregt, Sten and Middelburg, Luke M and El Mansouri, Brahim and van Zeijl, Henk W and May, Alexander and Erlbacher, Tobias and Zhang, Guoqi and Sarro, Pasqualina M},
  journal={IEEE Transactions on Electron Devices},
  volume={69},
  number={1},
  pages={4--10},
  year={2021},
  publisher={IEEE}
}

@Article{Romijn2022,
author={Romijn, Joost
and Vollebregt, Sten
and Middelburg, Luke M.
and Mansouri, Brahim El
and van Zeijl, Henk W.
and May, Alexander
and Erlbacher, Tobias
and Leijtens, Johan
and Zhang, Guoqi
and Sarro, Pasqualina M.},
title={Integrated 64 pixel {UV} image sensor and readout in a Silicon Carbide {CMOS} technology},
journal={Microsystems {\&} Nanoengineering},
year={2022},
month={Oct},
day={25},
volume={8},
number={1},
pages={114},
url={https://doi.org/10.1038/s41378-022-00446-3}
}

@INPROCEEDINGS{woodend,
  author={Woodend, L. J. and Gammon, P. M. and Shah, V. A. and Pérez-Tomás, A. and Li, F. and Hamilton, D. P. and Myronov, M. and Mawby, P. A.},
  booktitle={2016 European Conference on Silicon Carbide \& Related Materials (ECSCRM)}, 
  title={{C}ryogenic characterisation and modelling of commercial {S}i{C} {MOSFET}s}, 
  year={2016},
  volume={},
  number={},
  pages={1-1},
  doi={10.4028/www.scientific.net/MSF.897.557}}

@ARTICLE{elze,
  author = { Elzerman, J. M. and Hanson, R. and Willems van Beveren, L. H. and Witkamp, B. and Vandersypen, L. M. K. and Kouwenhoven, L. P.},
  title = {Single-shot read-out of an individual electron spin in a quantum dot},
  journal = {Nature},
  year = {2004},
  volume = {430},
  pages = {431}
}

@article{mfgz21,
	Author = {Gonzalez-Zalba, M. F. and de Franceschi, S. and Charbon, E. and Meunier, T. and Vinet, M. and Dzurak, A. S.},
	Da = {2021/12/01},
	Doi = {10.1038/s41928-021-00681-y},
	Id = {Gonzalez-Zalba2021},
	Isbn = {2520-1131},
	Journal = {Nature Electronics},
	Number = {12},
	Pages = {872--884},
	Title = {Scaling silicon-based quantum computing using {CMOS} technology},
	Ty = {JOUR},
	Url = {https://doi.org/10.1038/s41928-021-00681-y},
	Volume = {4},
	Year = {2021}}

@article{Burkard_Rev,
  title = {Semiconductor spin qubits},
  author = {Burkard, Guido and Ladd, Thaddeus D. and Pan, Andrew and Nichol, John M. and Petta, Jason R.},
  journal = {Rev. Mod. Phys.},
  volume = {95},
  issue = {2},
  pages = {025003},
  numpages = {58},
  year = {2023},
  month = {Jun},
  publisher = {American Physical Society},
  doi = {10.1103/RevModPhys.95.025003},
  url = {https://link.aps.org/doi/10.1103/RevModPhys.95.025003}
}

@misc{banerjee2024materialsquantumtechnologiesroadmap,
      title={Materials for Quantum Technologies: a Roadmap for Spin and Topology}, 
      author={N. Banerjee and C. Bell and C. Ciccarelli and T. Hesjedal and F. Johnson and H. Kurebayashi and T. A. Moore and C. Moutafis and H. L. Stern and I. J. Vera-Marun and J. Wade and C. Barton and M. R. Connolly and N. J. Curson and K. Fallon and A. J. Fisher and D. A. Gangloff and W. Griggs and E. Linfield and C. H. Marrows and A. Rossi and F. Schindler and J. Smith and T. Thomson and O. Kazakova},
      year={2024},
      eprint={2406.07720},
      archivePrefix={arXiv},
      primaryClass={cond-mat.mes-hall},
      url={https://arxiv.org/abs/2406.07720}, 
}

@Article{Awschalom2018,
author={Awschalom, David D.
and Hanson, Ronald
and Wrachtrup, J{\"o}rg
and Zhou, Brian B.},
title={Quantum technologies with optically interfaced solid-state spins},
journal={Nature Photonics},
year={2018},
month={Sep},
day={01},
volume={12},
number={9},
pages={516-527},
issn={1749-4893},
doi={10.1038/s41566-018-0232-2},
url={https://doi.org/10.1038/s41566-018-0232-2}
}

@article{Lohrmann_2017,
doi = {10.1088/1361-6633/aa5171},
url = {https://dx.doi.org/10.1088/1361-6633/aa5171},
year = {2017},
month = {jan},
publisher = {IOP Publishing},
volume = {80},
number = {3},
pages = {034502},
author = {Lohrmann, A and Johnson, B C and McCallum, J C and Castelletto, S},
title = {A review on single photon sources in Silicon Carbide},
journal = {Reports on Progress in Physics},
}

@article{abraham2021,
  title = {Nanotesla Magnetometry with the Silicon Vacancy in Silicon Carbide},
  author = {Abraham, John B. S. and Gutgsell, Cameron and Todorovski, Dalibor and Sperling, Scott and Epstein, Jacob E. and Tien-Street, Brian S. and Sweeney, Timothy M. and Wathen, Jeremiah J. and Pogue, Elizabeth A. and Brereton, Peter G. and McQueen, Tyrel M. and Frey, Wesley and Clader, B. D. and Osiander, Robert},
  journal = {Phys. Rev. Appl.},
  volume = {15},
  issue = {6},
  pages = {064022},
  numpages = {9},
  year = {2021},
  month = {Jun},
  publisher = {American Physical Society},
  doi = {10.1103/PhysRevApplied.15.064022},
  url = {https://link.aps.org/doi/10.1103/PhysRevApplied.15.064022}
}

@article{son_rev,
    author = {Son, Nguyen T. and Anderson, Christopher P. and Bourassa, Alexandre and Miao, Kevin C. and Babin, Charles and Widmann, Matthias and Niethammer, Matthias and Ul Hassan, Jawad and Morioka, Naoya and Ivanov, Ivan G. and Kaiser, Florian and Wrachtrup, Joerg and Awschalom, David D.},
    title = {Developing silicon Carbide for quantum spintronics},
    journal = {Applied Physics Letters},
    volume = {116},
    number = {19},
    pages = {190501},
    year = {2020},
    month = {05},
    issn = {0003-6951},
    doi = {10.1063/5.0004454},
    url = {https://doi.org/10.1063/5.0004454}
}

@article{lelis2014basic,
  title={Basic mechanisms of threshold-voltage instability and implications for reliability testing of {S}i{C} {MOSFET}s},
  author={Lelis, Aivars J and Green, Ron and Habersat, Daniel B and El, Mooro},
  journal={IEEE Transactions on Electron Devices},
  volume={62},
  number={2},
  pages={316--323},
  year={2014},
  publisher={IEEE}
}

@article{LELIS201832,
title = {{S}i{C} {MOSFET} threshold-stability issues},
journal = {Materials Science in Semiconductor Processing},
volume = {78},
pages = {32-37},
year = {2018},
note = {Wide band gap semiconductors technology for next generation of energy efficient power electronics},
issn = {1369-8001},
doi = {https://doi.org/10.1016/j.mssp.2017.11.028},
url = {https://www.sciencedirect.com/science/article/pii/S1369800117319807},
author = {Aivars J. Lelis and Ronald Green and Daniel B. Habersat},

}

@article{eastoe2024method,
  title={Method for efficient large-scale cryogenic characterization of {CMOS} technologies},
  author={Eastoe, Jonathan and Noah, Grayson M and Dutta, Debargha and Rossi, Alessandro and Fletcher, Jonathan D and Gomez-Saiz, Alberto},
  journal={IEEE Transactions on Instrumentation and Measurement},
   volume={74},
  number={},
  pages={1-10},
  doi={10.1109/TIM.2024.3497143},
  year={2024},
  publisher={IEEE}
}

@article{patra2017cryo,
  title={Cryo-{CMOS} circuits and systems for quantum computing applications},
  author={Patra, Bishnu and Incandela, Rosario M and Van Dijk, Jeroen PG and Homulle, Harald AR and Song, Lin and Shahmohammadi, Mina and Staszewski, Robert Bogdan and Vladimirescu, Andrei and Babaie, Masoud and Sebastiano, Fabio and others},
  journal={IEEE Journal of Solid-State Circuits},
  volume={53},
  number={1},
  pages={309--321},
  year={2017},
  publisher={IEEE}
}

@Article{Angus2007,
author={Angus, Susan J.
and Ferguson, Andrew J.
and Dzurak, Andrew S.
and Clark, Robert G.},
title={Gate-Defined Quantum Dots in Intrinsic Silicon},
journal={Nano Letters},
year={2007},
month={Jul},
day={01},
publisher={American Chemical Society},
volume={7},
number={7},
pages={2051-2055},
issn={1530-6984},
doi={10.1021/nl070949k},
url={https://doi.org/10.1021/nl070949k}
}

@article{beckers2019theoretical,
  title={Theoretical limit of low temperature subthreshold swing in field-effect transistors},
  author={Beckers, Arnout and Jazaeri, Farzan and Enz, Christian},
  journal={IEEE Electron Device Letters},
  volume={41},
  number={2},
  pages={276--279},
  year={2019},
  publisher={IEEE}
}

@book{neamen,
  author    = {D. Neamen},
  title     = {Semiconductor Physics and Devices},
  publisher = {McGraw-Hill},
  year      = {2012},
  edition   = {4th},
}

@book{sze,
  author    = {S.M. Sze and K.K. Ng},
  title     = {Physics of Semiconductor Devices},
  publisher = {Wiley},
  year      = {2007},
  edition   = {3rd},
}

@article{yoshi,
    author = {Yoshioka, Hironori and Senzaki, Junji and Shimozato, Atsushi and Tanaka, Yasunori  and Okumura, Hajime},
    title = {N-channel field-effect mobility inversely proportional to the interface state density at the conduction band edges of {S}i{O}2/4{H}-{S}i{C} interfaces},
    journal = {AIP Advances},
    volume = {5},
    number = {1},
    pages = {017109},
    year = {2015},
    month = {01},
    issn = {2158-3226},
    doi = {10.1063/1.4905781},
    url = {https://doi.org/10.1063/1.4905781}
}

@inproceedings{gui2018characterization,
  title={Characterization of 1.2 k{V} {S}i{C} power {MOSFET}s at cryogenic temperatures},
  author={Gui, Handong and Ren, Ren and Zhang, Zheyu and Chen, Ruirui and Niu, Jiahao and Wang, Fred and Tolbert, Leon M and Blalock, Benjamin J and Costinett, Daniel J and Choi, Benjamin B},
  booktitle={2018 IEEE Energy Conversion Congress and Exposition (ECCE)},
  pages={7010--7015},
  year={2018},
  organization={IEEE}
}

@inproceedings{chen2015,
author = {Chen, Han and Gammon, Peter Michael and Shah, Vishal A. and Fisher, C.A. and Chan, Chun W. and Jahdi, Saeed and Hamilton, D.P. and Jennings, Michael R. and Myronov, Maksym and Leadley, David R. and Mawby, Philip  Andrew},
title = {Cryogenic Characterization of Commercial {S}i{C} Power {MOSFET}s},
year = {2015},
month = {7},
volume = {821},
pages = {777--780},
booktitle = {Silicon Carbide and Related Materials 2014},
series = {Materials Science Forum},
publisher = {Trans Tech Publications Ltd},
doi = {10.4028/www.scientific.net/MSF.821-823.777},
}

@article{MATOCHA20081631,
title = {Challenges in {S}i{C} power {MOSFET} design},
journal = {Solid-State Electronics},
volume = {52},
number = {10},
pages = {1631-1635},
year = {2008},
note = {Papers Selected from the International Semiconductor Device Research Symposium 2007 – ISDRS 2007},
issn = {0038-1101},
doi = {https://doi.org/10.1016/j.sse.2008.06.034},
url = {https://www.sciencedirect.com/science/article/pii/S0038110108002001},
author = {Kevin Matocha},
}

@article{shen2022time,
  title={Time-dependent degradation mechanism of 1.2 k{V} {S}i{C} {MOSFET} under long-term high-temperature gate bias stress},
  author={Shen, Yutong and He, Zhiyuan and Shi, Yijun and Niu, Hao and Chen, Yuan and Liu, Chang and Chen, Yiqiang and Cai, Zongqi and Lu, Guoguang and Dai, Xianying},
  journal={IEEE Transactions on Electron Devices},
  volume={70},
  number={3},
  pages={1162--1167},
  year={2022},
  publisher={IEEE}
}

@article{gammon2013modelling,
  title={{M}odelling the inhomogeneous {S}i{C} {S}chottky interface},
  author={Gammon, PM and P{\'e}rez-Tom{\'a}s, Amador and Shah, VA and Vavasour, O and Donchev, E and Pang, JS and Myronov, Maksym and Fisher, Craig A and Jennings, MR and Leadley, David R and others},
  journal={Journal of Applied Physics},
  volume={114},
  number={22},
  year={2013},
  publisher={AIP Publishing}
}

@article{koliakoudakis2008cr,
  title={{C}r/4{H}-{S}i{C} {S}chottky contacts investigated by electrical and photoelectron spectroscopy techniques},
  author={Koliakoudakis, C and Dontas, J and Karakalos, S and Kayambaki, M and Ladas, S and Konstantinidis, G and Zekentes, K and Kennou, S},
  journal={physica status solidi (a)},
  volume={205},
  number={11},
  pages={2536--2540},
  year={2008},
  publisher={Wiley Online Library}
}

@article{pascu2020ultrashallow,
  title={Ultrashallow defects in {S}i{C} {MOS} capacitors},
  author={Pascu, Razvan},
  journal={Solid State Electronics Letters},
  volume={2},
  pages={79--84},
  year={2020},
  publisher={Elsevier}
}

@inproceedings{oka2023milli,
  title={Milli-kelvin analysis revealing the role of band-edge states in cryogenic {MOSFET}s},
  author={Oka, Hiroshi and Asai, Hidehiro and Inaba, Takumi and Shitakata, Shunsuke and Yui, Hitoshi and Fuketa, Hiroshi and Iizuka, Shota and Kato, Kimihiko and Nakayama, Takashi and Mori, Takahiro},
  booktitle={2023 International Electron Devices Meeting (IEDM)},
  pages={1--4},
  year={2023},
  organization={IEEE}
}

@article{roche2012tunable,
  title={A tunable, dual mode field-effect or single electron transistor},
  author={Roche, Beno{\i}t and Voisin, Benoit and Jehl, Xavier and Wacquez, Romain and Sanquer, Marc and Vinet, Maud and Deshpande, Veeresh and Previtali, Bernard},
  journal={Applied Physics Letters},
  volume={100},
  number={3},
  year={2012},
  publisher={AIP Publishing}
}

@article{son1995electron,
  title={Electron effective masses in 4H SiC},
  author={Son, NT and Chen, WM and Kordina, O and Konstantinov, AO and Monemar, B and Janz{\'e}n, E and Hofman, DM and Volm, D and Drechsler, M and Meyer, BK},
  journal={Applied physics letters},
  volume={66},
  number={9},
  pages={1074--1076},
  year={1995},
  publisher={American Institute of Physics}
}

@misc{powell2026data,
  author       = {Powell, M. and Rossi, A. and McGeough, C. and Parry, E. and Zotov, A.},
  year         = {2026},
  title        = {Data for: {``Reproducibility and variability in commercial SiC MOSFETs at deep-cryogenic''}},
  howpublished = {\textit{University of Strathclyde} (available at: \url{https://pureportal.strath.ac.uk/en/datasets/data-for-reproducibility-and-variability-in-commercial-sic-mosfet/})}
}

\clearpage
\onecolumngrid               % full‑width single column  :contentReference[oaicite:0]{index=0}

\appendix*                   % “single appendix” form   :contentReference[oaicite:1]{index=1}
\section{Extended datasets}   % bold, centred automatically
\label{sec:histo}

% ----- First wide figure -----
\begin{figure}[ht!]
  \centering
  \includegraphics[width=0.8\linewidth]{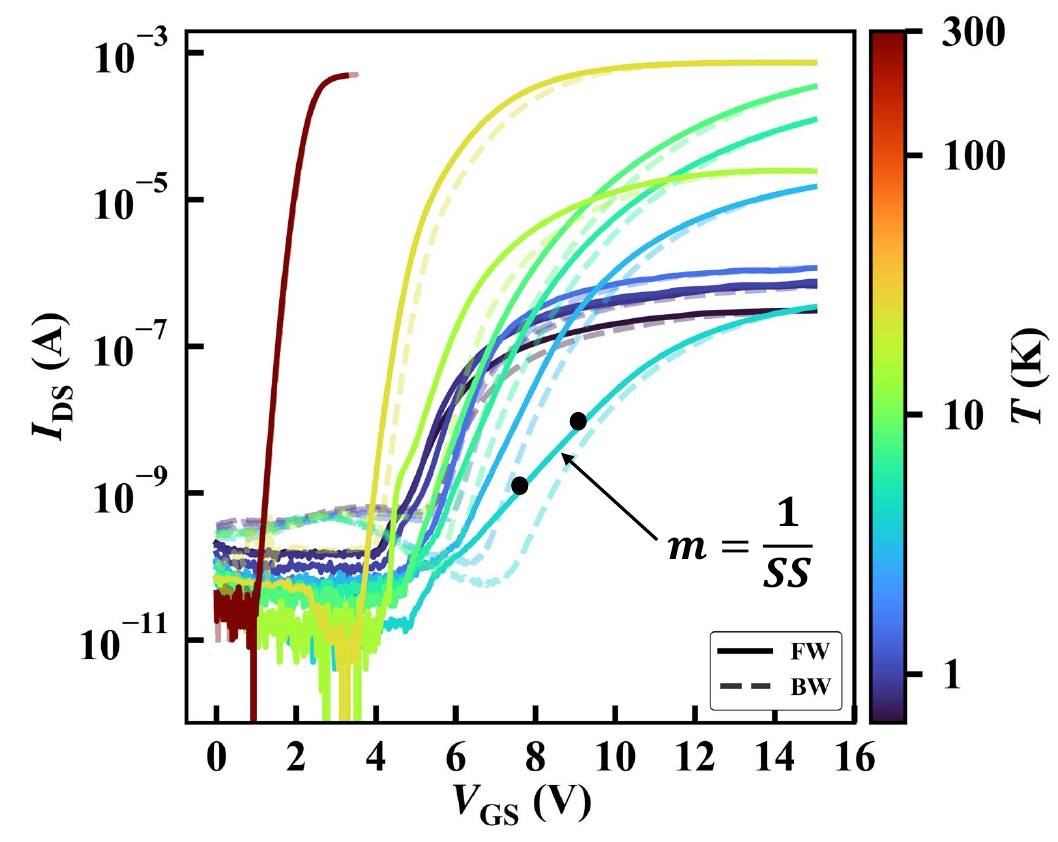}
  \caption{$I_\textup{DS}$-$V_\textup{GS}$ transfer characteristics for Device A as a function of temperature measured with forward (solid lines) and backward (dashed lines) gate sweeps. \( V_{\text{DS}}^\textup{0.65~K} \) = 3.8 V, \( V_{\text{DS}}^\textup{0.85~K} \) = 3.5 V, \( V_{\text{DS}}^\textup{1~K} \) = 3.2 V, \( V_{\text{DS}}^\textup{1.5~K} \) = 2.7 V, \( V_{\text{DS}}^\textup{3~K} \) = 2.4 V, \( V_{\text{DS}}^\textup{4~K} \) = 2.1 V, \( V_{\text{DS}}^\textup{6~K} \) = 2.4 V, \( V_{\text{DS}}^\textup{8~K} \) = 2.4 V, \( V_{\text{DS}}^\textup{15~K} \) = 2.0 V, \( V_{\text{DS}}^\textup{25~K} \) = 1.7 V, and \( V_{\text{DS}}^\textup{300~K} \) = 0.05 V. The fixed current values at which $SS$ is calculated (reciprocal of the gradient of the dotted line) are indicated by black dots.}
  \label{fig:AllTempTransfer}
\end{figure}

 \begin{figure*}[t]
  \centering
  \includegraphics[width=0.8\linewidth]{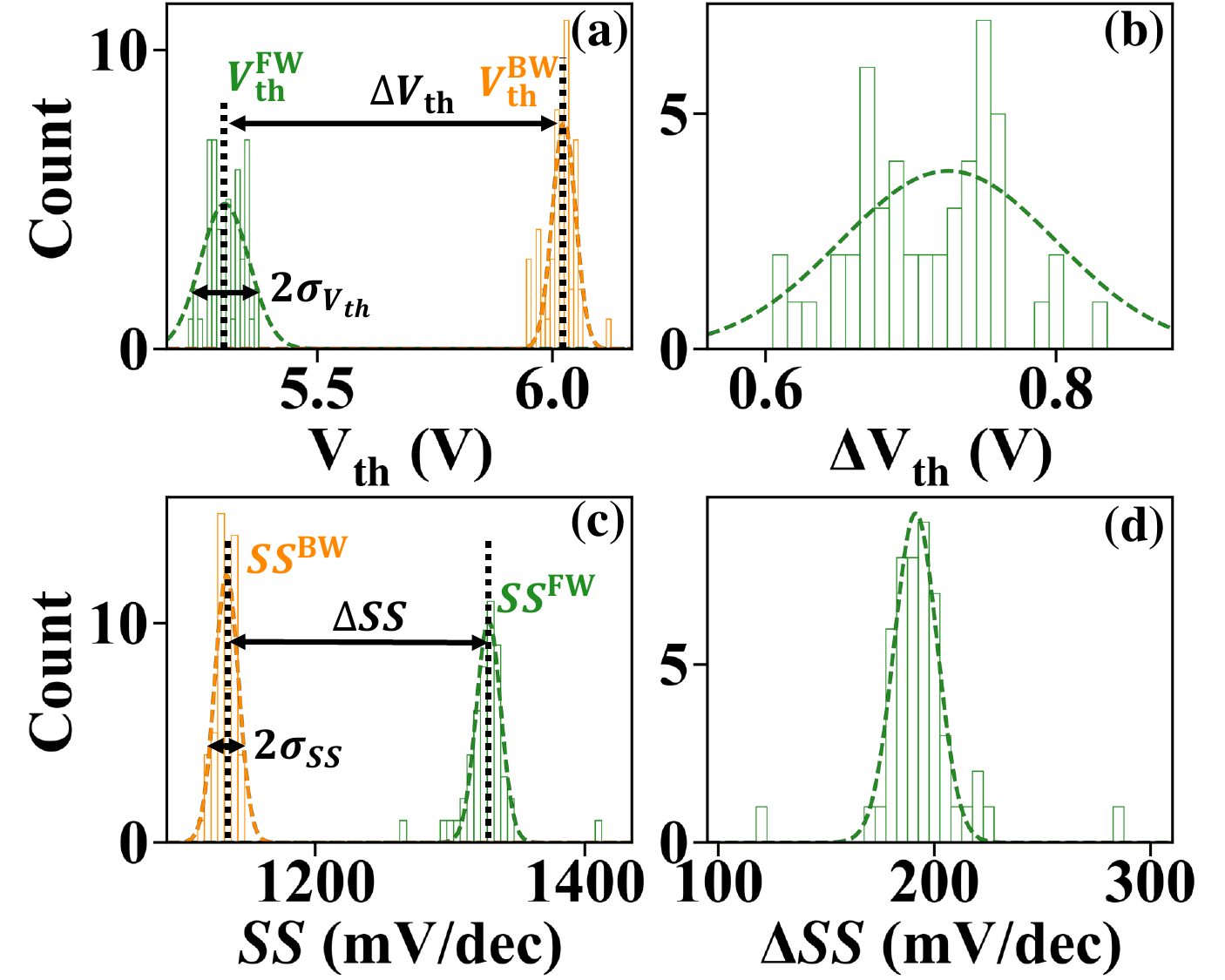}
  \caption{Device A's statistical distributions at $T=0.65~$K used to extract (a) $V_\textup{th}$ and $\sigma_{V_\textup{th}}$, (b) $\Delta V_{\textup{th}}$, (c) $SS$ and $\sigma_{\textup{SS}}$, (d) $\Delta SS$.}
  \label{fig:Histogram}
\end{figure*}

\begin{figure*}[t]
\centering
\includegraphics[scale=0.5]{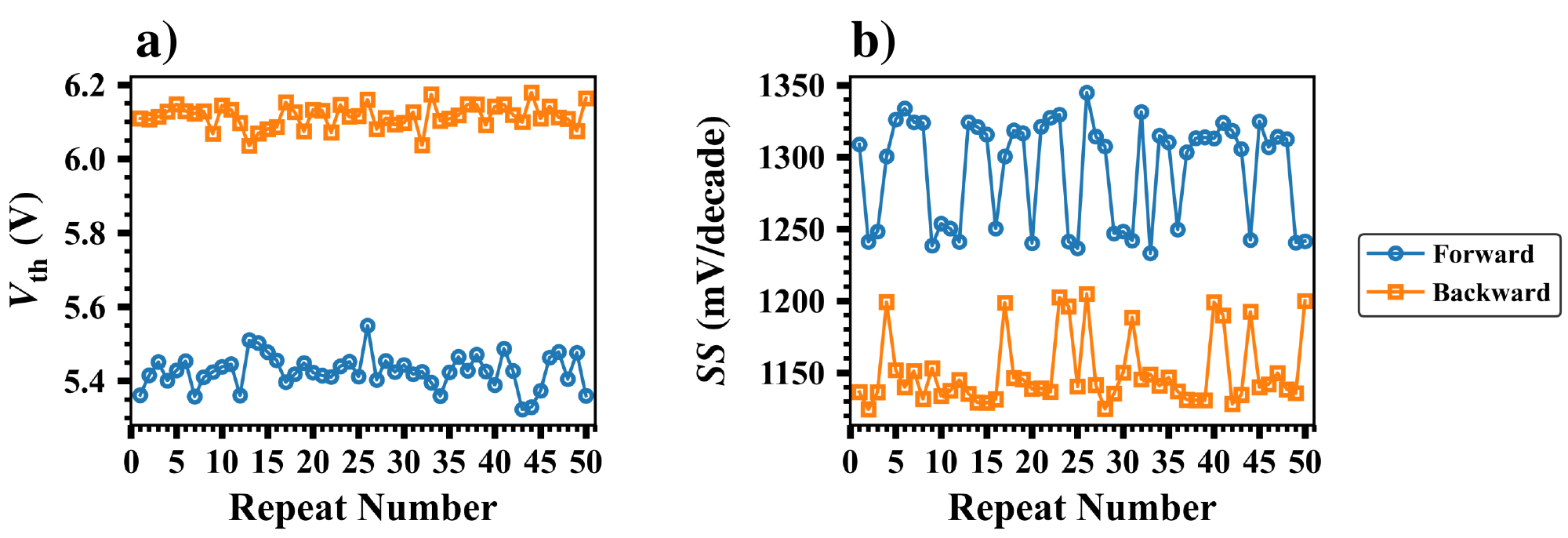}
\caption{Device A's distribution of extracted parameters from 50 repeated transfer characteristics measured at $T=0.65~$K and $V_\textup{DS}=4.9~$V after completion of the training procedure. Variations of (a) $V_\textup{th}$ and (b) $SS$ across repeat index, illustrating cycle-to-cycle variability. The absence of a dependence of parameter values on measurement runs indicates that the device behaviour is not subject to obvious degradation effects.}
\label{fig:Param_VS_Repeat}
\end{figure*}

\begin{figure*}[t]
\centering
\includegraphics[scale=0.4]{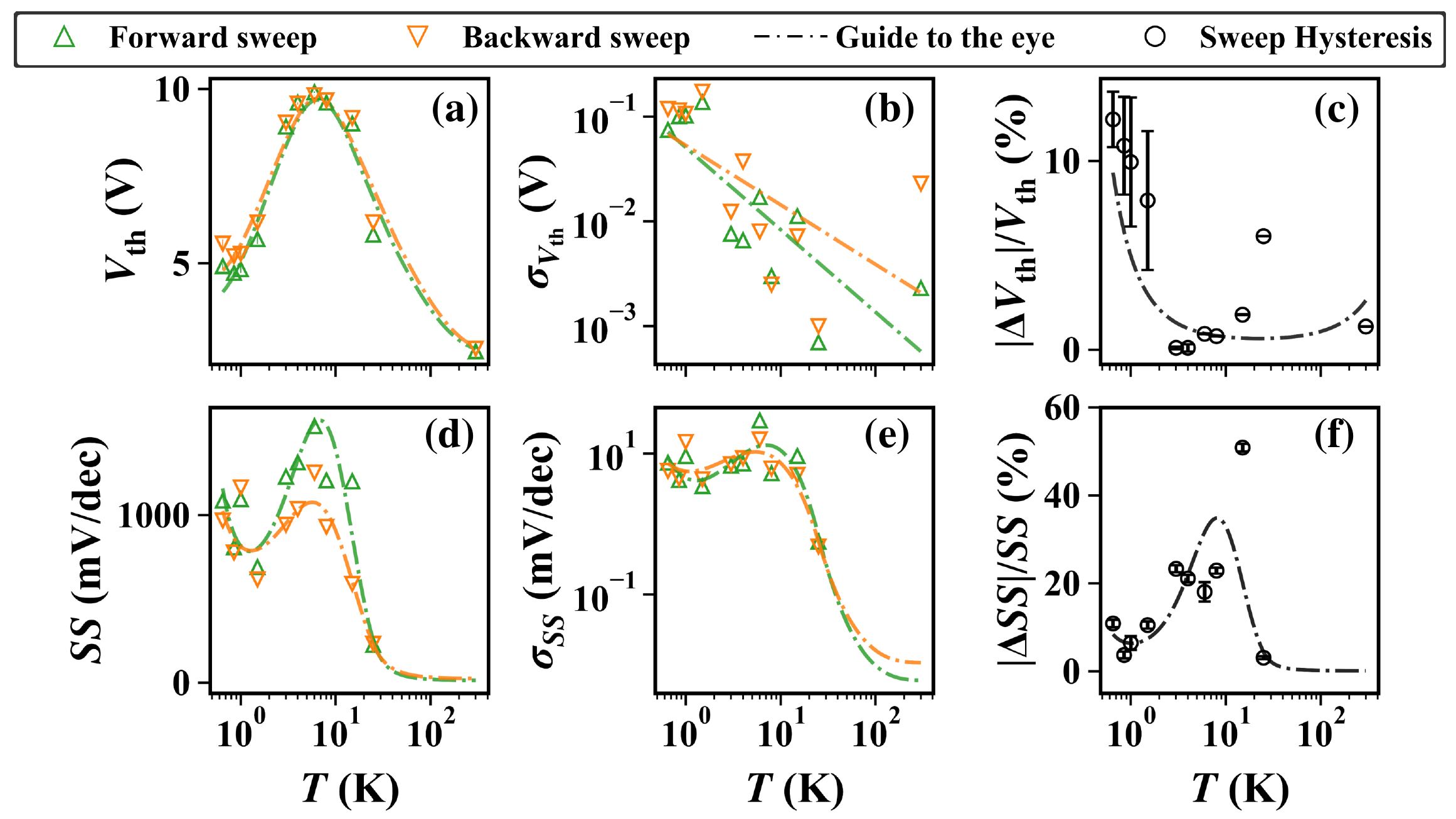}
\caption{Parameters extracted from statistical distributions of 50 characteristics as functions of temperature for Device B. The dot-dashed lines in each panel represent guides to the eye to aid evaluation of temperature dependencies. The error bars in (c) and (f) indicate a $\pm 1\sigma$ spread of the relevant distribution.}
\label{fig:Second_stat}
\end{figure*}

\begin{table}[htb]
\centering
\begin{tabular}{|c|c|c|}
\hline
$T$ (K) & $V_\textup{0}$ (V) & $d$ (nm) \\ \hline
0.65    & 1.07           & 23.67    \\ \hline
0.85    & 0.98           & 19.32    \\ \hline
1.5     & 1.64          & 13.46    \\ \hline
4       & 1.41           & 10.00       \\ \hline
6       & 1.24           & 8.83     \\ \hline
15      & 0.66           & 8.75     \\ \hline
\end{tabular}
\caption{The parameter results of fitting Device A's output characteristics to the Fowler-Nordheim tunneling model shown in Equation \ref{eq:FN_EQ}. }
\label{tab:my-table}
\end{table}

\begin{table}[h!]
\centering
\begin{tabular}{|c|c|c|c|c|c|c|}
\hline
$T$ (K) &
$\tau_1$ (s) &
\begin{tabular}[c]{@{}c@{}}$\tau_1$ Uncertainty\\ (s)\end{tabular} &
\begin{tabular}[c]{@{}c@{}}$\tau_1$ Relative Uncertainty\\ (\%)\end{tabular} &
$\tau_2$ (s) &
\begin{tabular}[c]{@{}c@{}}$\tau_2$ Uncertainty\\ (s)\end{tabular} &
\begin{tabular}[c]{@{}c@{}}$\tau_2$ Relative Uncertainty\\ (\%)\end{tabular} \\ \hline
8   & 989.9    & 9.8   & 0.99 & 17364.2 & 445 & 2.56 \\ \hline
15  & 2257.32  & 9.45  & 0.42 & 20207.8 & 459 & 2.27 \\ \hline
300 & 3401.02  & 15.8  & 0.46 & 23741.1 & 935 & 3.94 \\ \hline
\end{tabular}
\caption{Values and uncertainties of fitting parameters, $\tau_1$ and $\tau_2$, obtained from nonlinear least-squares fit of the time drift data reported in Fig.~\ref{fig:time}(a). The quoted uncertainties are the 1$\sigma$ standard deviations extracted from the fit covariance matrix. Relative uncertainties are given as percentages.}
\label{tab:tau_table}
\end{table}

\end{document}